\documentclass[graybox]{svmult}

\usepackage{mathptmx}       
\usepackage{helvet}         
\usepackage{courier}        
\usepackage{type1cm}        
\usepackage{makeidx}         
\usepackage{graphicx}        
\usepackage{multicol}        
\usepackage[bottom]{footmisc}

\makeindex             

\begin{document}

\title*{Low-Order Modelling of Blade-Induced Turbulence for RANS Actuator Disk Computations of Wind and Tidal Turbines}
\titlerunning{Low-Order Modelling of Blade-Induced Turbulence}
\author{Takafumi Nishino and Richard H. J. Willden}
\authorrunning{Nishino and Willden}
\institute{Takafumi Nishino \at Engineering Science, University of Oxford, \email{takafumi.nishino@eng.ox.ac.uk}
\and Richard H. J. Willden \at Engineering Science, University of Oxford \email{richard.willden@eng.ox.ac.uk}}
%
%
\maketitle

\abstract*{Modelling of turbine blade-induced turbulence (BIT) is discussed within the framework of three-dimensional Reynolds-averaged Navier-Stokes (RANS) actuator disk computations. We first propose a generic (baseline) BIT model, which is applied only to the actuator disk surface, does not include any model coefficients (other than those used in the original RANS turbulence model) and is expected to be valid in the limiting case where BIT is fully isotropic and in energy equilibrium. The baseline model is then combined with correction functions applied to the region behind the disk to account for the effect of rotor tip vortices causing a mismatch of Reynolds shear stress between short- and long-time averaged flow fields. Results are compared with wake measurements of a two-bladed wind turbine model of Medici and Alfredsson [Wind Energy, Vol. 9, 2006, pp. 219-236] to demonstrate the capability of the new model.} 

\abstract{Modelling of turbine blade-induced turbulence (BIT) is discussed within the framework of three-dimensional Reynolds-averaged Navier-Stokes (RANS) actuator disk computations. We first propose a generic (baseline) BIT model, which is applied only to the actuator disk surface, does not include any model coefficients (other than those used in the original RANS turbulence model) and is expected to be valid in the limiting case where BIT is fully isotropic and in energy equilibrium. The baseline model is then combined with correction functions applied to the region behind the disk to account for the effect of rotor tip vortices causing a mismatch of Reynolds shear stress between short- and long-time averaged flow fields. Results are compared with wake measurements of a two-bladed wind turbine model of Medici and Alfredsson [Wind Energy, Vol. 9, 2006, pp. 219-236] to demonstrate the capability of the new model.}

\section{Introduction}
\label{sec:1}
Modelling of turbulent mixing behind turbines is an important area of research for the wind and tidal power industries. Whilst recent eddy-resolving computations (such as large-eddy simulations and detached-eddy simulations) using a so-called actuator line method have given valuable insight into time-dependent features of wind/tidal turbine wakes (e.g., \cite{Ivanell_09}), it is still of great importance to develop a low-order model that satisfactorily predicts the characteristics of time-averaged flow around turbines at much lower computational cost. Such a low-order model would be useful for the design of wind and tidal power farms for the future.

The objective of the present study is to develop a reasonably generic low-order model that describes time-averaged flow around a horizontal-axis wind/tidal turbine of various designs and operating conditions. Our current model is based on the three-dimensional Reynolds-averaged Navier Stokes (RANS) equations coupled with an actuator disk model. To account for the effect of turbulence (or velocity fluctuations) induced by turbine blades---we refer to this as {\it blade-induced turbulence} (BIT) in this paper---we first introduce a simple baseline model, which requires the specification of BIT energy and scale but does not require any additional model coefficients (other than those used in the original RANS turbulence model). The baseline model is then combined with correction functions to account for the effect of rotor tip vortices. Details of the model are described below, followed by the results of a model validation study.

\section{Baseline Model}
\label{sec:2}

The baseline model used in this study is the one recently proposed by the authors \cite{Nishino_12}. The governing equations of the flow are the three-dimensional incompressible RANS equations, where the Reynolds stress tensor, $-\overline{u'_iu'_j}$, is modelled using the standard $k$-$\epsilon$ model of Launder and Spalding \cite{Launder_74}.

A simple actuator disk concept is used, where a turbine rotor is modelled as a stationary permeable disk of zero thickness placed perpendicular to the incoming flow. The effect of the rotor on the mean flow is considered as a loss of streamwise (or axial) momentum at the disk plane. The change in momentum flux (per unit disk-area and per unit fluid-density) is locally calculated as $S_U=\frac{1}{2}KU_d^2$, where $U_d$ is the local streamwise velocity at the disk plane and $K$ is a momentum loss factor (constant over the disk) to determine the thrust acting on the disk.

It is known that this actuator disk model (without taking account of BIT) yields much weaker turbulent mixing behind the disk compared to that measured behind a turbine rotor. In the baseline BIT model we assume that, at the disk plane, virtual turbine blades generate turbulence that is characterised by its turbulent kinetic energy $k_b$ and dissipation rate $\epsilon_b$. Two physical parameters are introduced to determine the values of $k_b$ and $\epsilon_b$: (i) the ratio of the energy converted to BIT to that removed from the mean flow at the disk plane, $\beta$, and (ii) a representative length scale for BIT, $l_b$. Note that $\beta$ and $l_b$ are model variables rather than model coefficients, as they depend on actual turbine design and operating conditions. $k_b$ is then calculated (locally over the disk plane) as $k_b=\beta S_U=\frac{1}{2}\beta KU_d^2$, whereas $\epsilon_b$ is estimated (based on the high Reynolds number equilibrium hypothesis) as $\epsilon_b=C_{\mu}^{\frac{3}{4}}k_b^{\frac{3}{2}}/l_b$, where $C_{\mu}=0.09$ (following the standard $k$-$\epsilon$ model).

To account for the combined effect of BIT and the turbulence coming from upstream of the turbine, we further assume that: (i) the disk plane is a special internal boundary that does not allow the transport of $k$ or $\epsilon$ via their diffusion (so that the transport of $k$ and $\epsilon$ through the disk plane is only via the convection from upstream to downstream) and (ii) at the disk plane, the BIT of $k_b$ and $\epsilon_b$ is mixed (in the time-averaged sense) with the {\it ambient} (or upstream) turbulence of $k_a$ and $\epsilon_a$, resulting in the {\it mixed} turbulence of $k_m$ and $\epsilon_m$ just downstream of the disk. Here $k_m$ is calculated as $k_m=k_a+k_b$, whereas $\epsilon_m$ is estimated by assuming the conservation of the time-integral of linearly decaying turbulent kinetic energy, i.e., $k_m\tau_m=k_a\tau_a+k_b\tau_b$, where $\tau_a=k_a/\epsilon_a$, $\tau_b=k_b/\epsilon_b$ and $\tau_m=k_m/\epsilon_m$ represent the initial eddy turnover time or lifetime for the ambient, blade-induced and mixed turbulence, respectively \cite{Nishino_12}. Eventually, the changes in $k$ and $\epsilon$ to be added to their transport equations at the disk plane (per unit disk-area) are calculated as
\begin{equation}
S_k=U_d(k_m-k_a)=U_dk_b\;,
\end{equation}
\begin{equation}
S_{\epsilon}=U_d(\epsilon_m-\epsilon_a)=U_d\left[\frac{(k_a+k_b)^2}{(k_a^2/\epsilon_a)+(k_b^2/\epsilon_b)}-\epsilon_a\right]\;.
\end{equation}

Generally, the two model variables, $\beta$ and $l_b$, should depend on the actual turbine design and may be given as functions of the distance, $r$, from the disk axis. In the present study, however, uniform values of $\beta$ and $l_b$ are given either to the entire disk surface or to the disk edge region defined by $(0.5d-w_{\rm{edge}})\leq r \leq 0.5d$, where $d$ is the disk diameter and $w_{\rm{edge}}$ ($= 0.1d$ in this study) is the width of the disk edge region. For the latter case, $\beta = 0$ is given to the rest of the disk surface.

\section{Tip Vortex Correction}
\label{sec:3}

As will be shown later, the baseline model tends to yield too strong/fast turbulent mixing in the core region ($r/d < 0.4$) in the near wake, even when BIT is given only around the disk edge. This suggests that further modifications are required on the modelling of turbulent mixing in the near wake region.

An important issue to be considered here is how large the Reynolds shear stress is (relative to the turbulent kinetic energy) behind a turbine, especially in the region where strong rotor tip vortices exist. It is known that for many two-dimensional turbulent shear flows where the production and dissipation of $k$ are close to equilibrium, the value of $\overline{u'_xu'_y}/k$ is around 0.3 (when the mean shear ${\rm d}U/{\rm d}y < 0$) or $-0.3$ (when ${\rm d}U/{\rm d}y > 0$) and this is the basis of the turbulent viscosity constant $C_{\mu} = (0.3)^2 = 0.09$ used in the standard $k$-$\epsilon$ model (see, e.g., Pope \cite{Pope_00}). In the mean shear layer behind a turbine rotor, however, experimental results have shown that the magnitude of $\overline{u'_xu'_y}/k$ can be significantly smaller than 0.3 \cite{Medici_11}. This is most likely due to the effect of tip vortices causing a mismatch between short-time (or phase) averaged and long-time averaged wake shear profiles. Specifically, tip vortices periodically create ``adverse'' velocity gradient regions, $\partial \langle u\rangle /\partial r < 0$ (where $\langle\phi\rangle$ denotes a short-time average of $\phi$ over a time scale relevant to the passing-through time of each tip vortex), and hence ``adverse'' Reynolds shear stress regions, $\langle u'_xu'_r\rangle>0$, whilst the long-time averaged wake shear direction is $\partial U/\partial r > 0$ and therefore $\overline{u'_xu'_r}<0$. This explains why the long-time-averaged wake shear behind a rotor does not diffuse quickly despite its high level of turbulent kinetic energy.

To account for the above effect of tip vortices within the framework of RANS actuator disk computations, we consider applying empirical correction functions to the so-called tip vortex region. The streamwise extent of the region will be given as a model variable, $l_{\rm tip}$, which may be further modelled as a function of turbine design and operating conditions in future studies. For the present study, we introduce three correction functions, $f_\mu$, $f_k$ and $f_{\epsilon 1}$, into the standard $k$-$\epsilon$ model:
\begin{equation}
\frac{Dk}{Dt}=\frac{\partial}{\partial x_j}\left( \frac{\nu_t}{\sigma_k}\frac{\partial k}{\partial x_j} \right)-\overline{u'_iu'_j}\frac{\partial U_i}{\partial x_j}f_k-\epsilon\;,
\end{equation}
\begin{equation}
\frac{D\epsilon}{Dt}=\frac{\partial}{\partial x_j}\left( \frac{\nu_t}{\sigma_\epsilon}\frac{\partial \epsilon}{\partial x_j} \right)-\overline{u'_iu'_j}\frac{\partial U_i}{\partial x_j}f_{\epsilon 1}C_{\epsilon 1}\frac{\epsilon}{k}-C_{\epsilon 2}\frac{\epsilon^2}{k}\;,
\end{equation}
\begin{equation}
-\overline{u'_iu'_j}=\nu_t\left(\frac{\partial U_i}{\partial x_j}+\frac{\partial U_j}{\partial x_i} \right)-\frac{2}{3}k\delta_{ij}\;,\;\;\;\;
\nu_t=f_\mu C_\mu\frac{k^2}{\epsilon}\;,
\end{equation}
where $C_\mu = 0.09$, $C_{\epsilon 1} = 1.44$, $C_{\epsilon 2} = 1.92$, $\sigma_k = 1.0$ and $\sigma_\epsilon = 1.3$ (following the standard $k$-$\epsilon$ model). To reduce the eddy viscosity and thus mitigate turbulent mixing in the tip vortex region behind the disk (located at $x=0$), we model $f_\mu$ as follows:
\begin{equation}
f_\mu=1\;\;\;\;\; {\rm for}\;\;\;\;\;x\leq 0\;,\; l_{\rm tip}\leq x\;,
\end{equation}
\begin{equation}
f_\mu=1-f_r(1-f_{\mu {\rm min}})\;\;\;\;\; {\rm for}\;\;\;\;\;0\leq x\leq l^*_{\rm tip}\;,
\end{equation}
\begin{equation}
f_\mu=1-f_r(1-f_{\mu {\rm min}})\frac{1}{2}\left( 1+\cos\frac{\pi (x-l^*_{\rm tip})}{l_{\rm tip}-l^*_{\rm tip}} \right)\;\;\;\;\; {\rm for}\;\;\;\;\;l^*_{\rm tip}\leq x\leq l_{\rm tip}\;,
\end{equation}
where $f_{\mu {\rm min}}=0.1$ is given in this study, following experimental observations \cite{Medici_11} (note, however, that this is also a model variable and should generally depend on the rotor tip-speed ratio and the number of blades). $l^*_{\rm tip}$ is the streamwise extent of the region where $f_\mu$ does not change (corresponding to the region where tip vortices are stable); we assume $l^*_{\rm tip}=\frac{1}{2}l_{\rm tip}$ in the present model. $f_r$ is a damping (sinusoidal) function in the radial ($r$) direction, normalised such that $f_r=1$ at $r=0.5d$ (where the rotor edge is located) and monotonically decreases to $f_r=0$ at $r=0$ and $d$.

A major difficulty in this correction is that the change in $\nu_t$ due to the introduction of $f_\mu$ affects not only the strength of turbulent diffusion but also the production of $k$ and $\epsilon$. This necessitates the introduction of $f_k$ and $f_{\epsilon 1}$ in Eqs (3) and (4), respectively; however, we assume $f_{\epsilon 1}=1$ in this study for simplicity.\footnote{Here the introduction of $f_k$ is justified since, in the tip vortex region, the production of $k$ should be linked to the mean of the magnitude of short-time averaged velocity gradient and the corresponding Reynolds stress rather than those for the long-time averaged flow field, i.e., $f_k$ accounts for the difference between $-\overline{u'_iu'_j}\frac{\partial U_i}{\partial x_j}$ (known in the model) and $\overline{\arrowvert -\langle u'_iu'_j\rangle\frac{\partial \langle u_i \rangle}{\partial x_j}\arrowvert}$ (unknown in the model). Meanwhile, physical interpretation of $f_{\epsilon 1}$ seems less straightforward since the production term of $\epsilon$ includes $\epsilon$ and $k$, which should also be divided into two components corresponding to the short-time and long-time averaged flow fields, respectively. This seems far beyond the capability of single-scale eddy viscosity models (perhaps we require a proper multi-scale model for such discussion; see, e.g., Sagaut et al. \cite{Sagaut_06}) and hence we assume $f_{\epsilon 1} = 1$ in this study.} Preliminary computations with $f_k = 1$ yielded too small $k$ in the near wake, whereas those with $f_k = 1/f_\mu$ (i.e., fully cancelling out the effect of $f_\mu$ on the production of $k$) resulted in too large $k$ in the near wake. To maintain the level of turbulent kinetic energy in the near wake comparable to that measured in the experiments \cite{Medici_11} (and also to minimise its sensitivity to the value of $l_{\rm tip}$), we model $f_k$ as $f_k=1+\gamma_k(1-f_\mu)/f_\mu$ with the model coefficient $\gamma_k=0.85$ in this study (note that $\gamma_k = 0$ and 1 correspond to the two extreme cases, $f_k = 1$ and $1/f_\mu$, respectively).
\begin{table}[b]
\caption{Experimental conditions}
\label{tab:1}
\begin{tabular}{p{2cm}p{1.5cm}p{1.5cm}p{2cm}p{4.3cm}}
\hline\noalign{\smallskip}
$U_{\rm in}$ & TSR & $C_T$ & Yaw angle & FST intensity  \\
\noalign{\smallskip}\svhline\noalign{\smallskip}
8.14 [m/s] & 3.87 & 0.899 & 0 [deg.] & 4.5$\%$ ($x/d=0$), 2.5$\%$ ($x/d=9$) \\
\noalign{\smallskip}\hline\noalign{\smallskip}
\end{tabular}
\end{table}
\begin{figure}[b]
\sidecaption
\includegraphics[scale=.68,trim = 0 10 10 10]{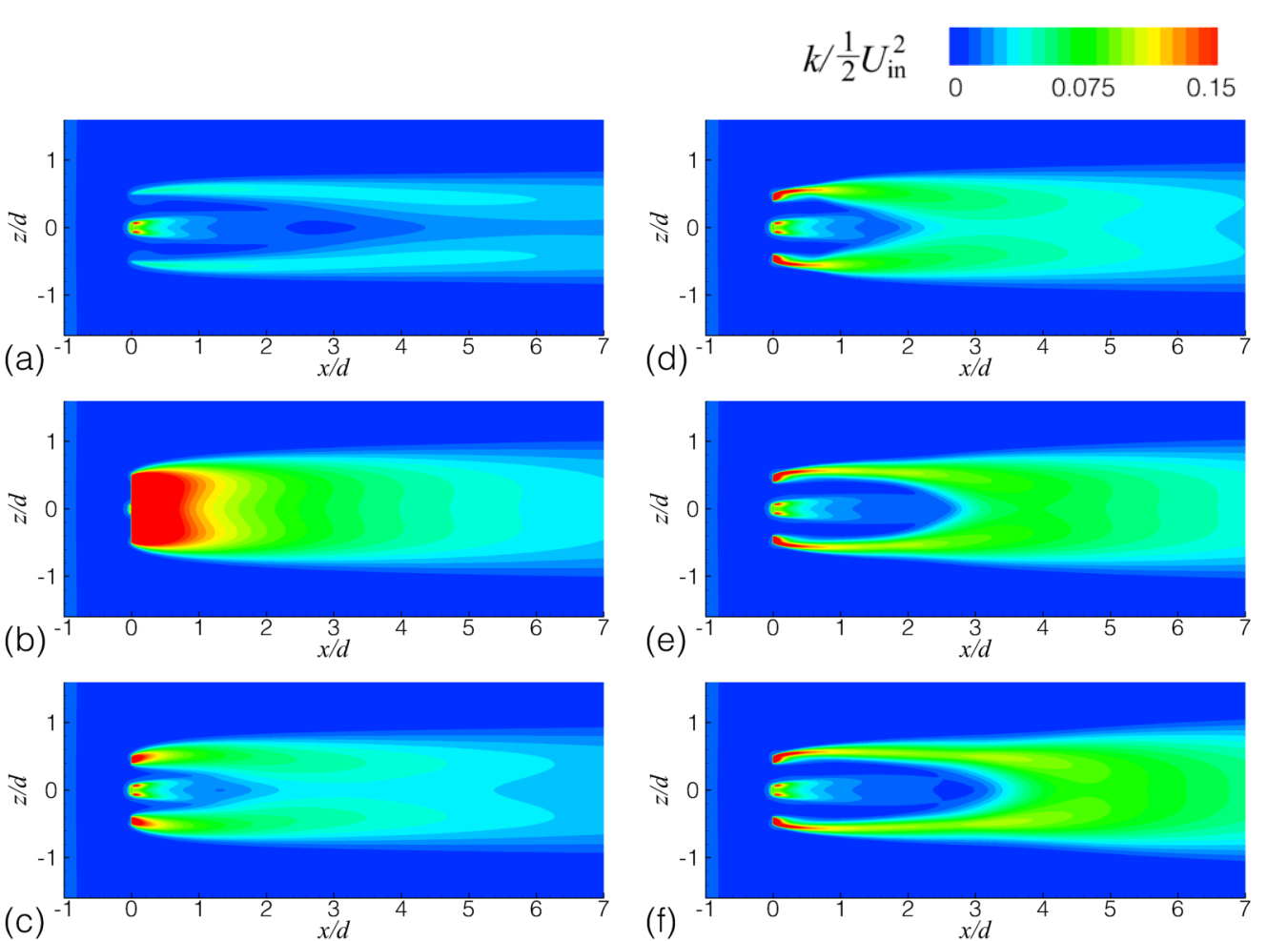}
\caption{Contours of turbulent kinetic energy across a horizontal plane at hub height: (a) without BIT, (b) baseline model (for the entire actuator disk), (c) baseline model (only for $0.4 < r/d < 0.5$), and (d-f) baseline model (only for $0.4 < r/d < 0.5$) with the tip vortex correction ($l_{\rm tip}/d = 1$, 3, 5 for d, e, f, respectively).}
\label{fig:1}
\end{figure}
\section{Model Validation}
\label{sec:4}

A model validation study was performed for wind tunnel tests of a two-bladed wind turbine model of Medici and Alfredsson \cite{Medici_06, Medici_11}. The tunnel test section is 1.2m wide and 0.8m high. The tunnel ceiling is adjustable, so that the freestream velocity, $U_{\rm in}$, is constant throughout the test section (when the turbine is not installed). The rotor diameter $d$ is 0.18m and its hub height from the floor is 0.24m. The hub diameter is 20mm. Table 1 summarises experimental conditions; for this particular set of experiments, a turbulence-generating grid was installed, providing freestream turbulence (FST) intensities of $4.5\%$ at $x/d = 0$ (turbine location) and $2.5\%$ at $x/d = 9$. Here the thrust coefficient $C_T$ and the tip-speed ratio (TSR) are based on the freestream velocity and therefore slightly different from those reported earlier \cite{Medici_06}.
 
For the computations, the turbine rotor was modelled as an actuator disk of zero thickness with a small non-permeable disk embedded at the centre to account for the effect of the rotor hub. The supporting tower of the turbine was not modelled. Computations were performed using FLUENT 12 (together with its User-Defined Function module). The solver is based on a finite volume method and is nominally 2nd-order accurate in space. All computations were performed as steady state.

Figures 1 and 2 present results for $C_T = 0.899$, $\beta = 0.30$ and $l_b = 0.1d$ (except for the case without BIT). The case without BIT yields too small turbulent kinetic energy and thus too slow wake recovery. The baseline model applied to the entire disk yields too large turbulent kinetic energy in the core region ($r/d < 0.4$) and hence too early wake recovery behind the disk. The baseline model applied only to the edge region still yields too strong turbulent mixing in the core region in the near wake, and the tip vortex correction provides better results in both near and far wake. Further validation studies are needed to examine the robustness of the model.

\begin{figure}[t]
\sidecaption
\includegraphics[scale=0.21,trim = 10 60 40 170,angle=270]{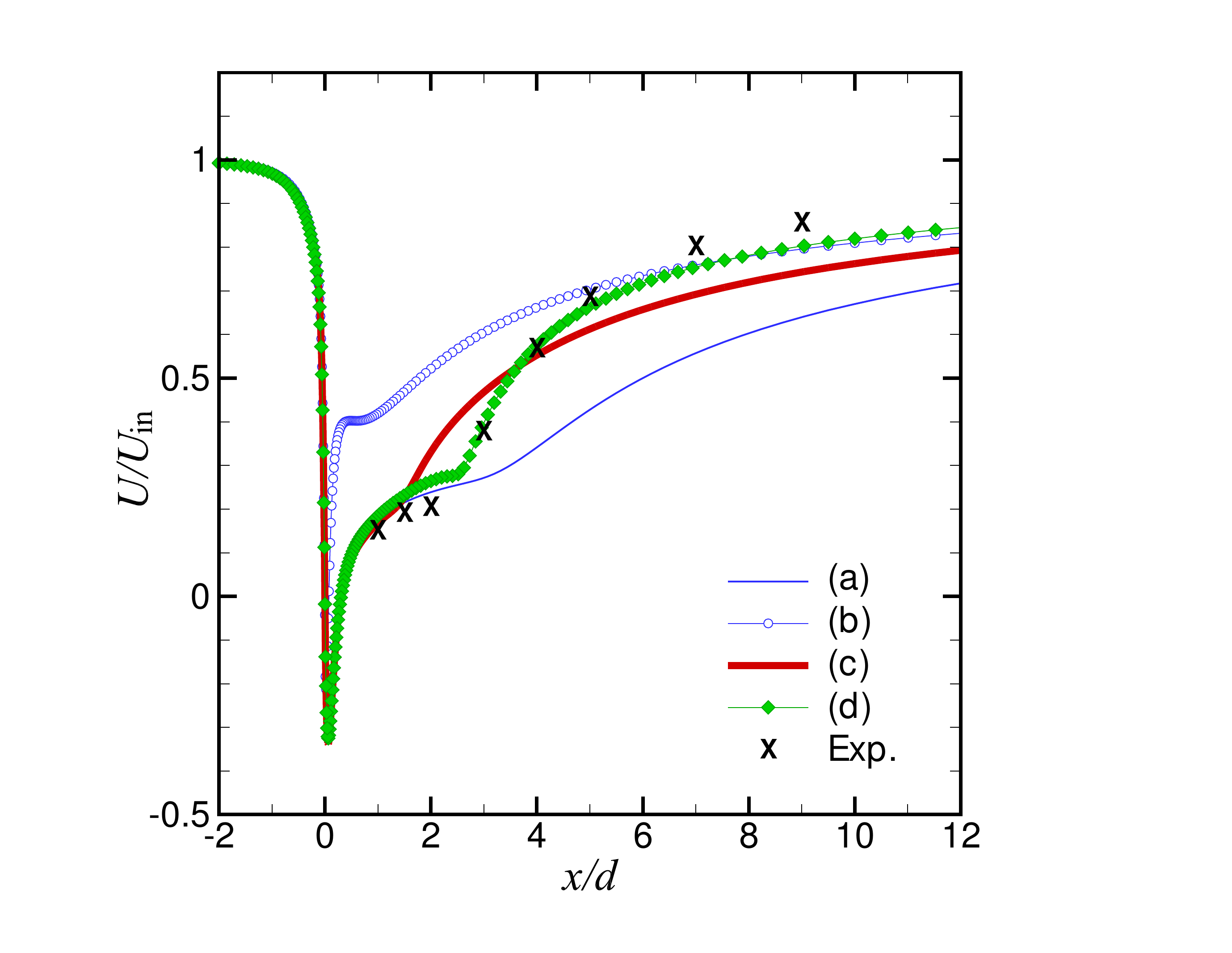}
\includegraphics[scale=0.21,trim = 10 70 40 240,angle=270]{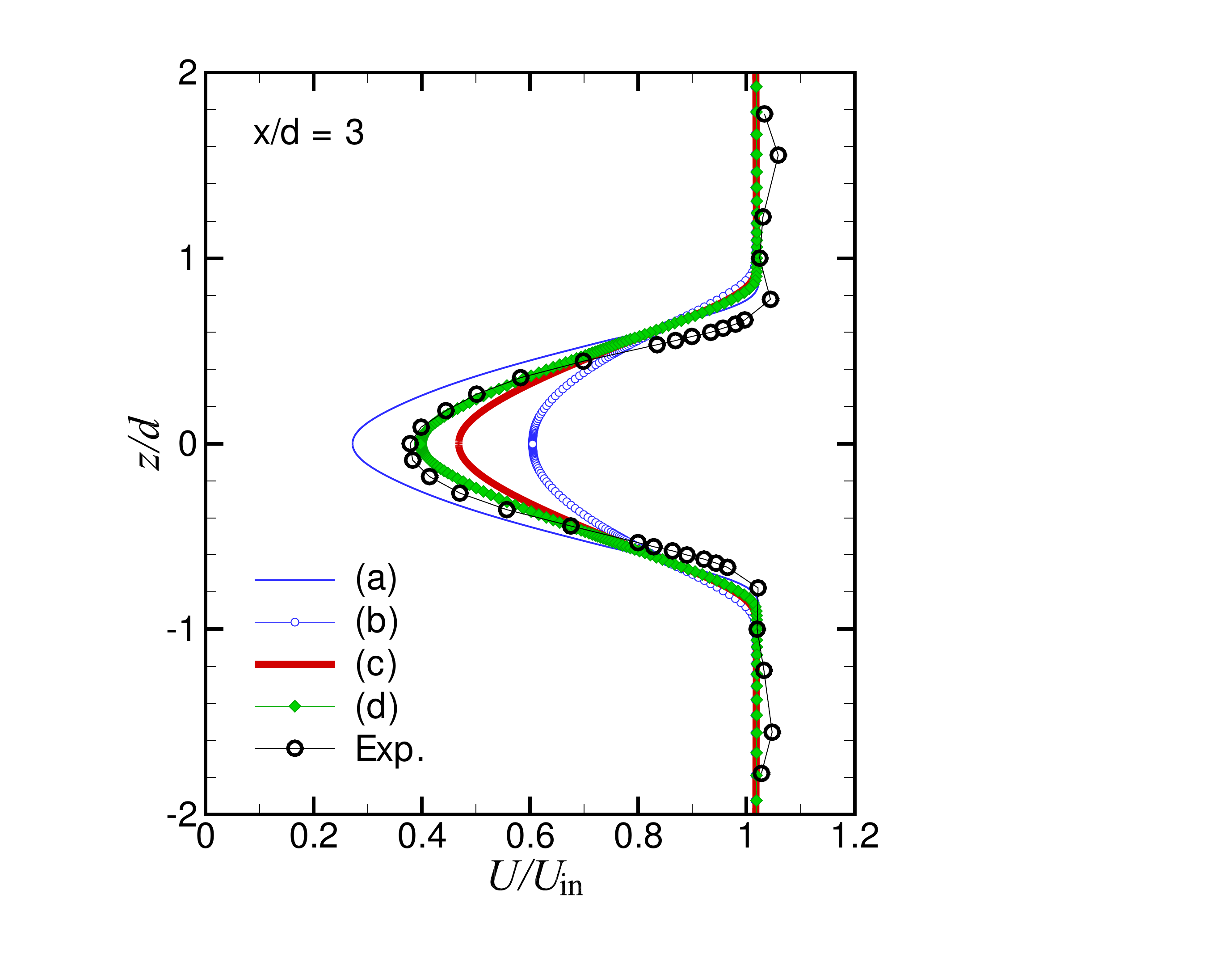}
\includegraphics[scale=0.21,trim = 10 70 40 240,angle=270]{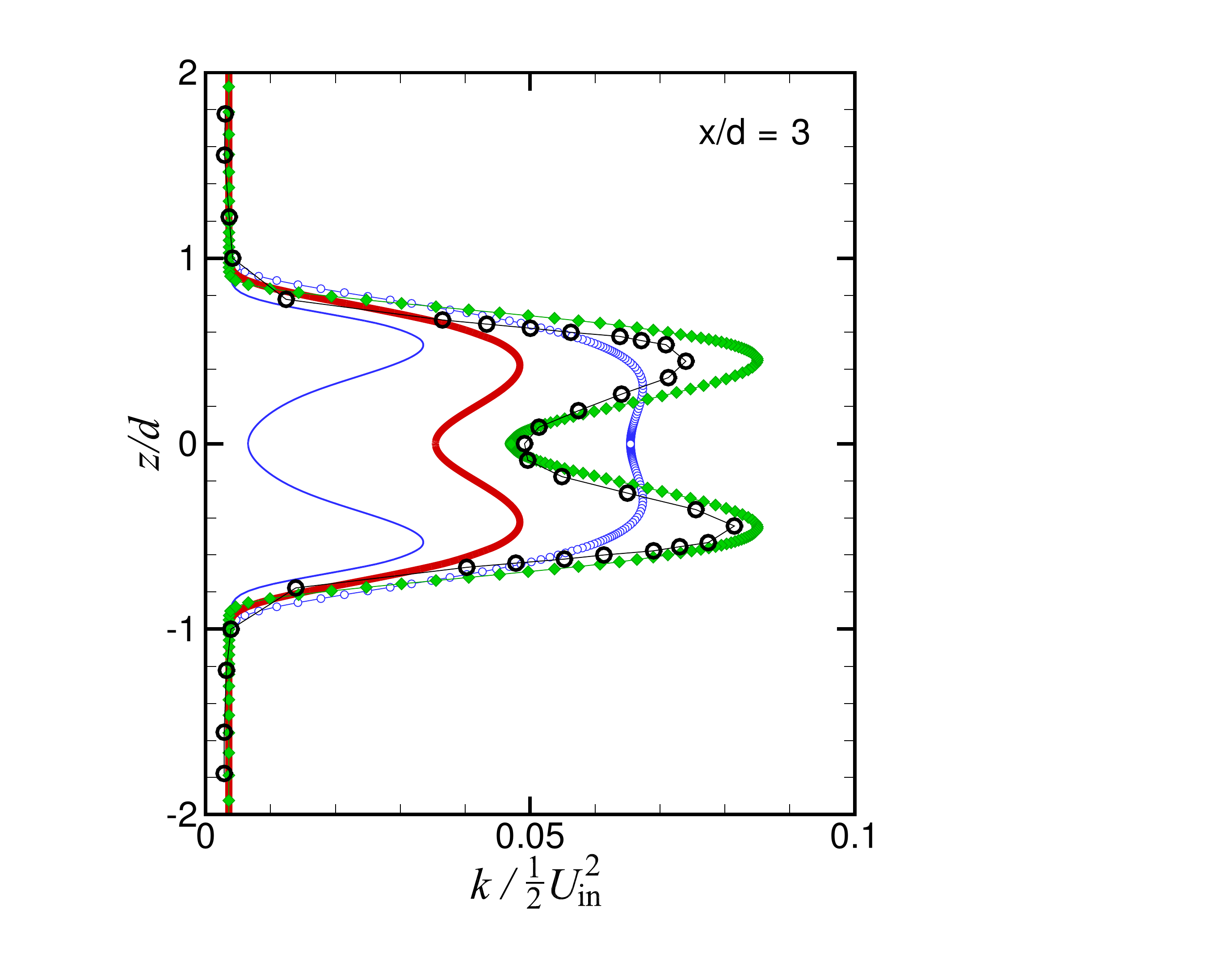}
\caption{Centreline streamwise velocity (left), streamwise velocity profiles at $x/d = 3$ (centre) and turbulent kinetic energy profiles at $x/d = 3$ (right): (a) without BIT, (b) baseline model (for the entire actuator disk), (c) baseline model (only for $0.4 < r/d < 0.5$), (d) baseline model (only for $0.4 < r/d < 0.5$) with the tip vortex correction ($l_{\rm tip}/d = 3$), and experimental data \cite{Medici_06}.}
\label{fig:2}
\end{figure}

\begin{acknowledgement}
The authors gratefully acknowledge the support of the Oxford Martin School, University of Oxford, who have funded this research. The authors would also like to thank Dr Davide Medici and Prof. Henrik Alfredsson for providing us with their experimental data.  
\end{acknowledgement}

%
%
%

\end{document}